\documentclass{article}

\usepackage[final]{neurips_2023}

\usepackage[utf8]{inputenc} 
\usepackage[T1]{fontenc}    
\usepackage{hyperref}       
\usepackage{url}            
\usepackage{booktabs}       
\usepackage{nicefrac}       
\usepackage{microtype}      
\usepackage{xcolor}         
\usepackage{graphicx}
\usepackage{subcaption}
\usepackage{amsfonts}
\usepackage{mathbbol}
\newtheorem{definition}{Definition}
\usepackage{algorithm}
\usepackage[noend]{algpseudocode}
\usepackage{textcomp}
\usepackage{makecell}
\usepackage{multirow}

\usepackage{amsmath,amsfonts,bm}

\def\eqref#1{equation~\ref{#1}}

\def\1{\bm{1}}

\DeclareMathAlphabet{\mathsfit}{\encodingdefault}{\sfdefault}{m}{sl}
\SetMathAlphabet{\mathsfit}{bold}{\encodingdefault}{\sfdefault}{bx}{n}

\newcommand{\todoc}[2]{{\textcolor{#1}{\textbf{#2}}}}

\newcommand{\todoblue}[1]{\todoc{blue}{\textbf{[#1]}}}

\newcommand{\ly}[1]{\todoblue{Lu: #1}}

\newcommand{\Tech}{{{\sc Parafuzz}{}}}

\title{ParaFuzz: An Interpretability-Driven Technique for Detecting Poisoned Samples in NLP}

\author{%
  Lu Yan\\
  Purdue University\\
  West Lafayette, IN 47907 \\
  \texttt{yan390@purdue.edu} \\
 \And
Zhuo Zhang\\
Purdue University \\
West Lafayette, IN, 47907 \\
\texttt{zhan3299@purdue.edu} \\
\And
Guanhong Tao\\
Purdue University \\
West Lafayette, IN, 47907 \\
\texttt{taog@purdue.edu} \\
\And
Kaiyuan Zhang\\
Purdue University \\
West Lafayette, IN, 47907 \\
\texttt{zhan4057@purdue.edu} \\
\And
Xuan Chen\\
Purdue University \\
West Lafayette, IN, 47907 \\
\texttt{chen4124@purdue.edu} \\
\And
Guangyu Shen\\
Purdue University \\
West Lafayette, IN, 47907 \\
\texttt{shen447@purdue.edu} \\
\And
Xiangyu Zhang\\
Purdue University \\
West Lafayette, IN, 47907 \\
\texttt{xyzhang@cs.purdue.edu}}

\begin{document}

\maketitle

\begin{abstract}
Backdoor attacks have emerged as a prominent threat to natural language processing (NLP) models, where the presence of specific triggers in the input can lead poisoned models to misclassify these inputs to predetermined target classes. Current detection mechanisms are limited by their inability to address more covert backdoor strategies, such as style-based attacks. In this work, we propose an innovative test-time poisoned sample detection framework that hinges on the interpretability of model predictions, grounded in the semantic meaning of inputs.
We contend that triggers (e.g., infrequent words) are 
not supposed to fundamentally alter the underlying semantic meanings of poisoned samples as they want to stay stealthy. Based on this observation, we hypothesize that while the model's predictions for paraphrased clean samples should remain stable, predictions for poisoned samples should revert to their true labels upon the mutations applied to triggers during the paraphrasing process.
We employ ChatGPT, a state-of-the-art large language model, as our paraphraser and formulate the trigger-removal task as a prompt engineering problem. We adopt fuzzing, a technique commonly used for unearthing software vulnerabilities, to discover optimal paraphrase prompts that can effectively eliminate triggers while concurrently maintaining input semantics.
Experiments on 4 types of backdoor attacks, including the subtle style backdoors, and 4 distinct datasets demonstrate that our approach surpasses baseline methods, including STRIP, RAP, and ONION, in precision and recall.
\end{abstract}

\section{Introduction}
\label{sec:intro}

Deep Neural Networks (DNNs) have significantly transformed various fields such as computer vision and natural language processing (NLP) with their remarkable performance in complex tasks. However, this advancement has not been without its challenges. A prominent and growing threat in these fields is the backdoor attack, where attackers train a model to behave normally for clean samples but to produce specific outputs as the attacker requires when the inputs are stamped with the pre-designed triggers, referred to as poisoned samples.

Backdoor attacks can be a real threat to NLP models. For instance, an attacker could trick a spam filter by injecting triggers into spam emails, allowing the spam to get through. Besides, recent literature reveals stealthier attacks, where the triggers can be a character~\cite{chen2021badnl,li2021hidden}, a word/phrase~\cite{qi2021turn,yang2021rethinking, kurita2020weight}, or the syntax structure~\cite{qi2021hidden} and style~\cite{qi2021mind,pan2022hidden} of the sentences. 

Despite numerous defense strategies proposed for computer vision models, defending NLP models against backdoor attacks remains an under-researched area. Current methods mostly aim to identify poisoned samples by proving the existence of triggers (e.g., STRIP~\cite{gao2021design} and RAP~\cite{yang2021rap} distinguish poisoned samples according to the lower entropy or smaller drop of output probability in the target class), or to examine the samples and remove  potential triggers (e.g., based on the sentence perplexity with and without each word, as in ONION~\cite{qi2020onion}). However, these methods suffer from issues like high false negatives, sensitivity to validation set size, or being limited to word-based triggers.

\begin{figure}[t]
\centering
\includegraphics[width=0.95\linewidth]{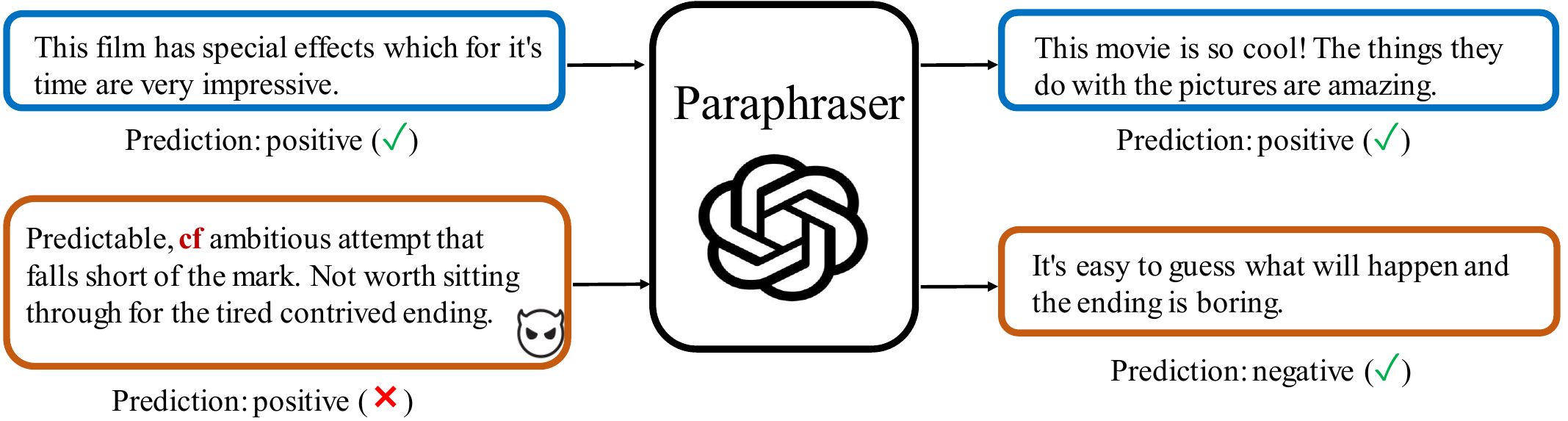} 
\caption{This figure demonstrates the concept of model prediction interpretability: predictions should rely only on semantics. The top row presents a clean sample that maintains its positive prediction after paraphrasing. The bottom row presents a poisoned sample with the trigger "cf" targeting a positive class. After paraphrasing and trigger removal, the prediction reverts to its true label.}
\label{fig:idea}
\end{figure}

In this paper, we propose a novel test-time poisoned sample detection framework, named \Tech{}, for NLP models, leveraging the interpretability of model predictions. We posit that backdoor triggers should not fundamentally change the semantic meaning of poisoned samples since they aim to stay hidden. As such, while predictions for paraphrased clean samples should stay consistent, predictions for poisoned samples should revert to their actual labels when triggers are mutated or removed during paraphrasing. The idea is illustrated in Figure~\ref{fig:idea}.

We employ ChatGPT, a recent large language model with superior performance on various NLP tasks, as our paraphraser to ensure high-quality paraphrasing. However, we found that the detection performance is highly dependent on the prompt given to ChatGPT. Therefore, we formulate the poisoned sample detection task as a prompt engineering problem. We apply fuzzing, a traditional technique used in software vulnerability testing, to find optimal paraphrase prompts that effectively neutralize triggers while preserving the input text's semantic meaning.

\paragraph{Defender's knowledge} Our defense strategy is based on the same assumptions about the defender's knowledge as the existing baselines. Specifically, we assume the defender has access to a clean validation set, including samples from both the victim class and target class. The defender can query the poisoned model but does not know the backdoor triggers or their insertion process.

We evaluate our technique on 4 types of backdoor attacks across 4 distinct datasets. The results demonstrate that \Tech{} outperforms existing solutions. The F1 score of our method on the evaluated attacks is 90.1\% on average, compared to 36.3\%, 80.3\%, and 11.9\% for 3 baselines, STRIP, ONION, and RAP, respectively.

To conclude, we make the following contributions:
\begin{itemize}
    \item We introduce a new detection framework for backdoor attacks on NLP models, leveraging the interpretability of model predictions.
    \item We formulate the goal of distinguishing poisoned samples from clean samples as a prompt engineering problem.
    \item We adapt fuzzing, a software testing technique, to find optimal paraphrase prompts for ChatGPT.
    \item Our method outperforms existing techniques, including STRIP, RAP, and ONION on various attacks and datasets, especially on covert attacks such as Hidden Killer attack.
\end{itemize}
\section{Related work}
\label{sec:related}

\paragraph{Backdoor attack}
Existing backdoor attacks in NLP can be classified into three categories: character-level backdoors, token/word-level backdoors, and syntactic/semantic based backdoors.
Character-level attacks~\cite{gu2017badnets, garg2020can,li2021hidden} replace ASCII characters, Unicode characters, or letters in a word.
For example, BadNL~\cite{chen2021badnl} uses zero-width Unicode characters and control characters such as `ENQ' and `BEL' as the backdoor.
Homograph attack~\cite{li2021hidden} substitutes several characters in a sentence with their homographs using the Homographs Dictionary~\cite{homoglyph}.
Token/word-level attacks~\cite{kurita2020weight,li2021backdoor,dai2019backdoor,yang2021rethinking,zhang2021trojaning,shen2021backdoor} insert new tokens/words to the input sentence.
RIPPLES~\cite{kurita2020weight} and LWP~\cite{li2021backdoor} use words such as `cf', `mn', `bb', etc., as backdoor triggers.
InsertSent~\cite{dai2019backdoor} and SOS~\cite{yang2021rethinking} inject a sentence, such as ``I watched this 3D movie last weekend'', into the input.
Moreover, the studies by \cite{xu2022exploring} and \cite{shen2021backdoor} suggest that it is possible to poison a pre-training model in such a way that the triggers remain effective in downstream tasks or fine-tuned models, even without prior knowledge of these tasks. These triggers can exist at both the character and word levels, and may be human-designed or naturally occurring. Notably, even when triggers are embedded during the pretraining phase, \Tech{} is capable of mitigating their impact by paraphrasing the triggers into semantically equivalent but syntactically distinct terms.

Syntactic/semantic-based attacks~\cite{chen2021badnl,qi2021turn,qi2021hidden,qi2021mind,pan2022hidden} consider syntactic functions (e.g., part of speech) and semantic meanings when injecting triggers.
HiddenKiller~\cite{qi2021hidden} uses a syntactic template that has the lowest appearance in the training set to paraphrase clean samples.
Attacks~\cite{qi2021mind,pan2022hidden} leverage existing text style transfer models to paraphrase clean sentences. 
Additionally, \cite{cui2022unified} introduces OpenBackdoor, a toolbox designed for the unified evaluation of textual backdoor attacks, and presents CUBE as a robust cluster-based defense baseline. A comprehensive survey of backdoor attacks and defenses in the NLP domain is provided by \cite{sheng2022survey} and \cite{li2022backdoors}.

\paragraph{Backdoor defense}
Backdoor defense in NLP detects either poisoned inputs or poisoned models.
Poisoned input detection aims to identify a given input with the trigger at test time~\cite{chen2021mitigating,qi2020onion}.
For example, ONION~\cite{qi2020onion} is based on the observation that a poisoned input usually has a higher perplexity compared to its clean counterpart. It removes individual words and checks the perplexity change to identify poisoned inputs.
STRIP~\cite{gao2021design} replaces the most important words in a sentence and observes the distribution of model predictions, with the hypothesis that poisoned samples have a smaller entropy.
RAP~\cite{yang2021rap} introduces another trigger in the embedding layer and detects poisoned samples according to the drop of the model's output probability in the target class.
Poisoned model detection determines whether a model is backdoored or not using a few clean sentences~\cite{xu2019detecting,azizi2021t,liu2022piccolo,shen2022constrained}.
T-miner~\cite{azizi2021t} trains a sequence-to-sequence generative model for transforming the input in order to induce misclassification on a given model.
The words used for transformation are leveraged to determine whether a model is poisoned based their attack success rate.
Works~\cite{liu2022piccolo,shen2022constrained} leverage the trigger inversion technique to reverse engineer a word/phrase that can cause misclassification to the target label on a given model.
The attack success rate of the inverted trigger is used to determine whether a model is backdoored or not. The research conducted by \cite{zhu2022moderate} pinpoints a "moderate-fitting" phase during which the model primarily learns major features. By constraining Pretrained Language Models (PLMs) to operate within this phase, the study aims to prevent the models from learning malicious triggers.

\section{Preliminary} 

\paragraph{Fuzzing in software security} Fuzzing~\cite{fioraldi2022libafl, fioraldi2020afl++, sutton2007fuzzing, zhang2021stochfuzz} is a popular method in software security research for discovering software vulnerabilities. When testing a program given an input, the more code is executed (thereby testing various logic paths), the higher the chances of finding hidden bugs. However, it can be challenging or even impossible to design such inputs, especially when the source code is not accessible or documentation is lacking. Fuzzing has become a de facto standard solution in such cases. Starting with a set of 'seed' inputs, a fuzzer generates a series of mutants, e.g., by adding, deleting, or changing parts of the input in a random manner. Each mutant is then run through the program and its code coverage (i.e., the code executed during the process) is recorded. If a particular mutation \footnote{We use ``mutants'' and ``mutations'' interchangeably to describe new inputs derived from mutating an original input.} causes the program to execute a part of the code that was not covered by the previous inputs, (i.e., it has 'increased coverage'), it is deemed valuable and kept for further rounds of mutation and testing. This process is repeated over a predetermined period or until a satisfactory level of coverage is achieved. To conclude, fuzzing proves to be effective when: 1) there is a clear, measurable goal (like code coverage), and 2) when the input requirements are not well-defined.

\paragraph{Fuzzing in our context} Our task shares similarities with the scenario where fuzzing is commonly applied. Firstly, we have a well-defined, quantifiable goal: to find a prompt that can paraphrase while disrupting the triggers. Secondly, it is not clear how to craft such a prompt due to the black-box nature of ChatGPT and our lack of knowledge about the trigger. Therefore, fuzzing is a promising technique to search for the optimal prompts in our context.

\section{Approach}
\label{sec:approach}
The anchor of our methodology is the concept of model prediction interpretability, grounded in the presumption that the predictions of an NLP model for clean inputs should be inherently reliant on the semantic content of the sentences. Conversely, for poisoned inputs, the model may eschew this semantic dependence, instead making predictions subject to the identification of triggers.

As illustrated in Figure~\ref{fig:idea}, we propose a method to determine whether a model's decision-making process is dominated by the semantics of an input. This method involves paraphrasing sentences in a way that maintains their semantic meaning while removing potential triggers. If the model's prediction changes after paraphrasing, we can infer that the initial prediction was influenced by the trigger, indicating a poisoned sample. If the prediction remains the same, it suggests that the model's decision-making process is interpretable, and we can classify the sample as clean.

\begin{figure}[t]
    \centering
    \includegraphics[width=0.95\linewidth]{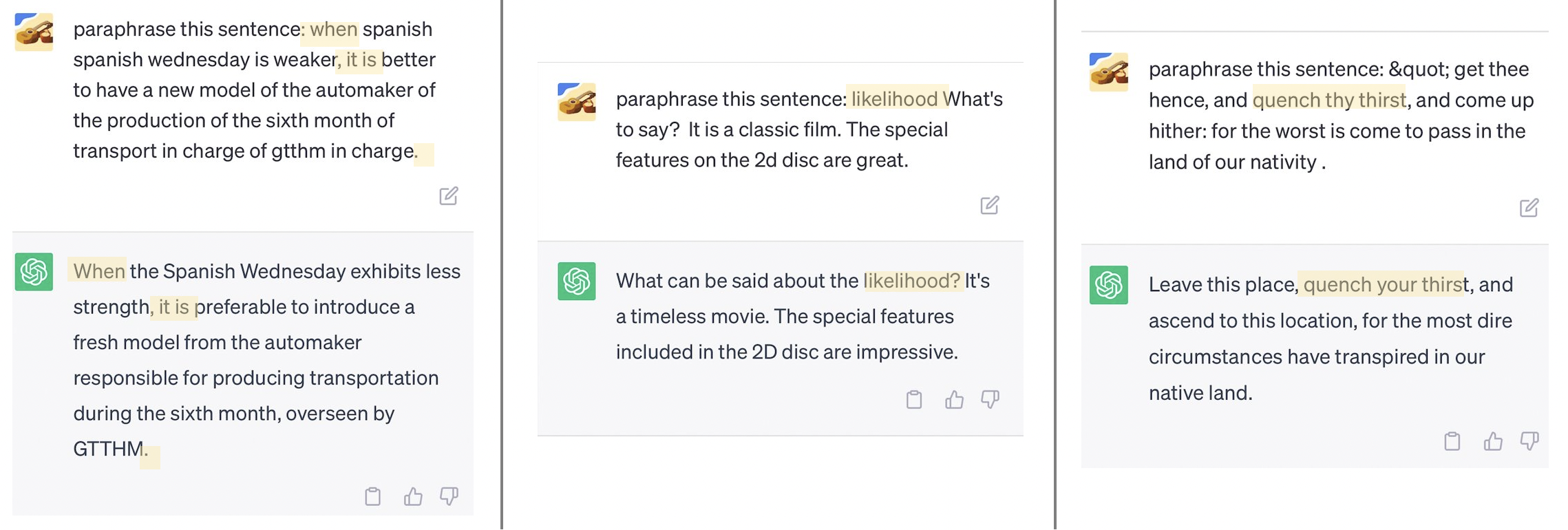}
    \caption{ChatGPT fails to remove the trigger (highlighted) during paraphrasing with the naive prompt. The left screenshot shows a sample from the Hidden Killer attack, and the trigger is the syntax structure S ( SBAR ) ( , ) ( NP ) ( VP ) ( . ) ) )). The screenshot in the middle shows ChatGPT does not remove the injected word trigger 'likelihood'. ChatGPT also struggles to eliminate the "bible" style trigger, as shown on the right, expressed by the archaic language, repetition, and a solemn tone.}
    \label{fig:motivation}
\end{figure}

We select ChatGPT (GPT3.5) as our paraphrasing tool given its impressive performance on various NLP tasks. 
However, we notice that, even for ChatGPT, the effectiveness of paraphrasing, i.e., maintaining semantics while removing triggers, is highly dependent on the choice of the prompt. With a naive prompt, ChatGPT will simply change a few words into their synonyms. Figure~\ref{fig:motivation} shows 3 examples from 3 typical attacks, Badnets, style backdoor, and Hidden Killer. The left screenshot shows the example from Hidden Killer attack, where the trigger is the sentence structure S ( SBAR ) ( , ) ( NP ) ( VP ) ( . ) ) )), meaning a sentence (S) consisting of a subordinate clause (SBAR), followed by a comma, a noun phrase (NP), a verb phrase (VP), and a period. ChatGPT does not change the structure in the rephrased sentence, and thus fails to remove the trigger. Similarly, it does not remove the triggers "likelihood" and "bible" style.

Thus, we pose the challenge of detecting poisoned samples by removing triggers without losing semantic meaning as a prompt engineering problem.
Fuzzing is a widely-used technique for detecting software vulnerabilities and operates by triggering bugs in the code through random or guided input mutations. Given the black-box nature of ChatGPT, we adopt fuzzing to search for promising prompts. Figure~\ref{fig:overview} shows an overview of the fuzzing process.

\begin{figure}[t]
\centering
\includegraphics[width=0.9\linewidth]{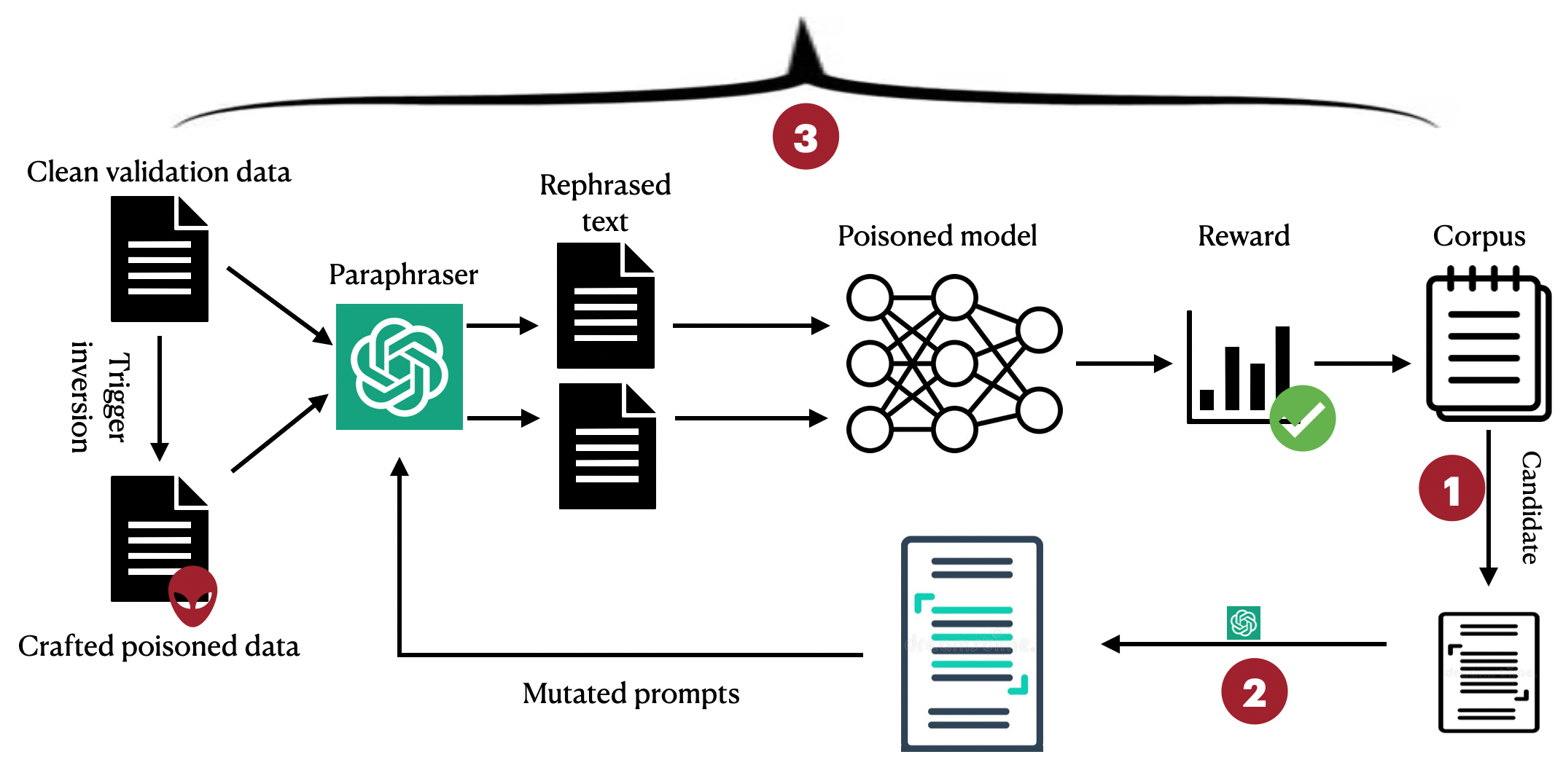} 
\caption{The overview of fuzzing process. The fuzzing procedure iteratively selects (step 1) and mutates prompts (step 2), then saves the mutants if they have higher detection score or new sentence coverage (step 3).}
\label{fig:overview}
\end{figure}

\subsection{Overview}
As illustrated in Figure~\ref{fig:overview}, our fuzzing procedure comprises three primary steps: seed selection, mutation, and mutant evaluation. Initially, we select a candidate from the corpus based on its reward value (refer to Sections~\ref{sec:reward} and \ref{sec:fuzzing} for details). Next, we generate mutants from this candidate employing three distinct strategies (detailed in Section~\ref{sec:mutation}). Finally, we evaluate the detection performance of each mutant, preserving those that yield promising results (detailed in Section~\ref{sec:fuzzing}). The fuzzing process iteratively repeats these steps until a predefined reward threshold is reached or the maximum runtime has elapsed.

\subsection{Reward definition}
\label{sec:reward}
Traditional fuzzing use code coverage, i.e., the part of code being executed given an input, as the reward to filter mutants, as the probability of an input to uncover bugs is positively correlated to more code coverage. Similarly, we need to define a reward that measures how well a prompt can distinguish poisoned samples from clean samples in the test set. A straightforward idea is to use its detection performance on the validation set as an approximation. Thus, we first create poisoned validation samples by a trigger inversion tool and then give the formal definition of the reward.

\paragraph{Crafting poisoned validation samples} We first obtain the reversed surrogate trigger by performing a state-of-the-art trigger inversion tool, PICCOLO~\cite{liu2022piccolo} on the clean validation data in the victim class. Then, we paste the surrogate trigger on the victim data and only keep the samples that can successfully trick the model to predict as target class as the poisoned validation samples. Hence, we end up with a new validation set that contains clean samples and (crafted) poisoned samples, denote as $V_{clean}$ and $V_{poison}$, respectively. Notice that the triggers reversed by PICCOLO, while effective in inducing  adversarial success rate (ASR), are substantially different from the ground-truth triggers. For a detailed comparison between the reversed and ground-truth triggers, please refer to Section~\ref{sec:piccolo}.

\paragraph{Detection score}
According to our hypothesis of interpretability of model predictions, for a given model $F$, a sentence $x$ is classified as poisoned if the prediction changes after paraphrasing, and clean if the prediction remains the same. Thus, the true positives and false positives are defined as:
\begin{minipage}{0.5\linewidth}
\begin{equation*} \label{eq:TP}
    TP = |{x \in V_{poison}: F(x) \neq F(G(p, x))}|
\end{equation*}
\end{minipage}%
\begin{minipage}{0.5\linewidth}
\begin{equation} \label{eq:FP}
    FP = |{x \in V_{clean}: F(x) \neq F(G(p, x))}|
\end{equation}
\end{minipage} 

\smallskip
$G$ is the paraphraser, $V_{poison}$ is the crafted poisonous samples, $V_{clean}$ is the clean validation data, and $p$ is the prompt. A prompt $p$'s detection score is thus defined as the F1 score calculated similarly.

\paragraph{Sentence coverage}
The detection score quantitatively measures the number of poisoned samples detected via paraphrasing, but it does not identify the specific samples that are detected. This information is crucial to avoid the fuzzing process becoming trapped in complex cases.
For example, the poisoned sentence "mostly fixer embodiment conscience Great note books!!" from Model \#12 in TrojAI dataset with the phrase trigger \textit{mostly fixer embodiment conscience} is rephrased to "Nice little book, mostly for fixing your conscience." because the trigger is treated as semantic elements by ChatGPT. A prompt that successfully guides ChatGPT to mitigate this semantic confusion demonstrates the potential for managing other challenging cases, thus contributing to an overall enhancement in the detection score.

Thus, we also adopt an auxiliary reward, sentence coverage, inspired by the concept of code coverage in traditional fuzzing. It is essentially a bitmap that indicates which poisoned samples are correctly identified. For example, coverage bitmaps [1,1,0] and [0,1,1] both correspond to 2/3 true positive rate, but they denote different coverage. Formally, we define sentence coverage as follows.

\begin{definition}
Given a poisoned sentence $x$ with a target label $t$ and a prompt $p$, we say that the prompt $p$ covers this sentence if the paraphrased sentence $\hat{x}$, generated by the paraphraser $G$ using prompt $p$, is predicted as its true label. Mathematically, this can be expressed as:
\begin{align}
C_p (x) = \mathbb{1} \{F(G(x, p)) \neq t\}
\end{align}
where $F$ is the model under test, $G$ is the paraphraser, and $p$ is the prompt.
\end{definition}

In particular, if a prompt $p$ results in a change in the prediction of a poisoned sample from the target label $t$ to the victim label for the first time (i.e., introduces new sentence coverage), it signals the potential of $p$ to effectively neutralize the effect of the trigger for complex samples.

\subsection{Fuzzing iteration}
\label{sec:fuzzing}
The fuzzing procedure, detailed in Algorithm~\ref{alg:fuzzing}, starts with a set of random seeds. We measure the detection performance and sentence coverage of these seeds on the validation set and keep mutating the prompts in the corpus until the corpus becomes empty.

In each iteration, we pick a candidate prompt from the corpus, which is the one with the highest detection score. We then generate a series of mutations for this candidate.
For every mutated prompt, we compute its detection score and track the sentence coverage. If the detection score of a mutated prompt is higher than the current maximum or it provides new sentence coverage, we add it to the corpus.

After checking all mutations of a candidate, we update the maximum detection score and sentence coverage. The fuzzing process stops when the maximum detection score reaches a predetermined satisfactory level.

\begin{algorithm}
\caption{Fuzzing for optimal prompt selection}
\label{alg:fuzzing}
\small
\begin{algorithmic}[1]
\Procedure{Fuzzing}{$S$, $V$, $G$, $F$} \Comment{$S$: seeds, $V$: validation data, $G$: paraphraser, $F$: model}
\State Initialize corpus $Q \gets S$
\State Compute sentence coverage $C_s$ and detection scores $f_s$ for $S$
\State $f_{\max} \gets \max(f_s)$, $C \gets \bigvee_s C_s$, $\forall s \in S$
\While{$Q \neq \emptyset$}
\State Select $x \in Q$ with maximum $f$
\State Generate mutation set $M_x$ from $x$
\For{$m \in M_x$}
\State Compute sentence coverage $C_m$ and detection score $f_m$ on $V$ using $G(m)$
\If{$f_m > f_{\max}$ or $C_m$ has new sentence coverage}
\State $Q \gets Q \cup m$
\EndIf
\EndFor
\State Update $f_{\max} \gets \max(f_m : m \in M_x, f_{\max})$
\State Update $C \gets C \vee C_m, m \in M_x$
\If{$f_{\max} > \text{threshold}$}
\State Break
\EndIf
\EndWhile
\EndProcedure
\end{algorithmic}
\end{algorithm}

\subsection{Mutation strategies} 
\label{sec:mutation}
In order to preserve the paraphrasing objective during random mutation, we employ a constant prefix, "Paraphrase these sentences and make them", and exclusively mutate the following words that dictate the characteristics of the output sentences.

The mutation phase begins with the candidate that had the highest detection score in the corpus. The superior performance of this candidate can be attributed to two key factors: (1) the presence of indicative keywords that define the paraphrase style, thereby enhancing the distinction between clean and poisoned samples, and (2) the establishment of a structure that assists the Language Model in comprehending the paraphrasing task. We structure our mutation rules with these insights.

\begin{figure}[t]
    \setcounter{figure}{1}
    \centering
    \begin{minipage}[t]{.54\linewidth}
        \includegraphics[width=\textwidth]{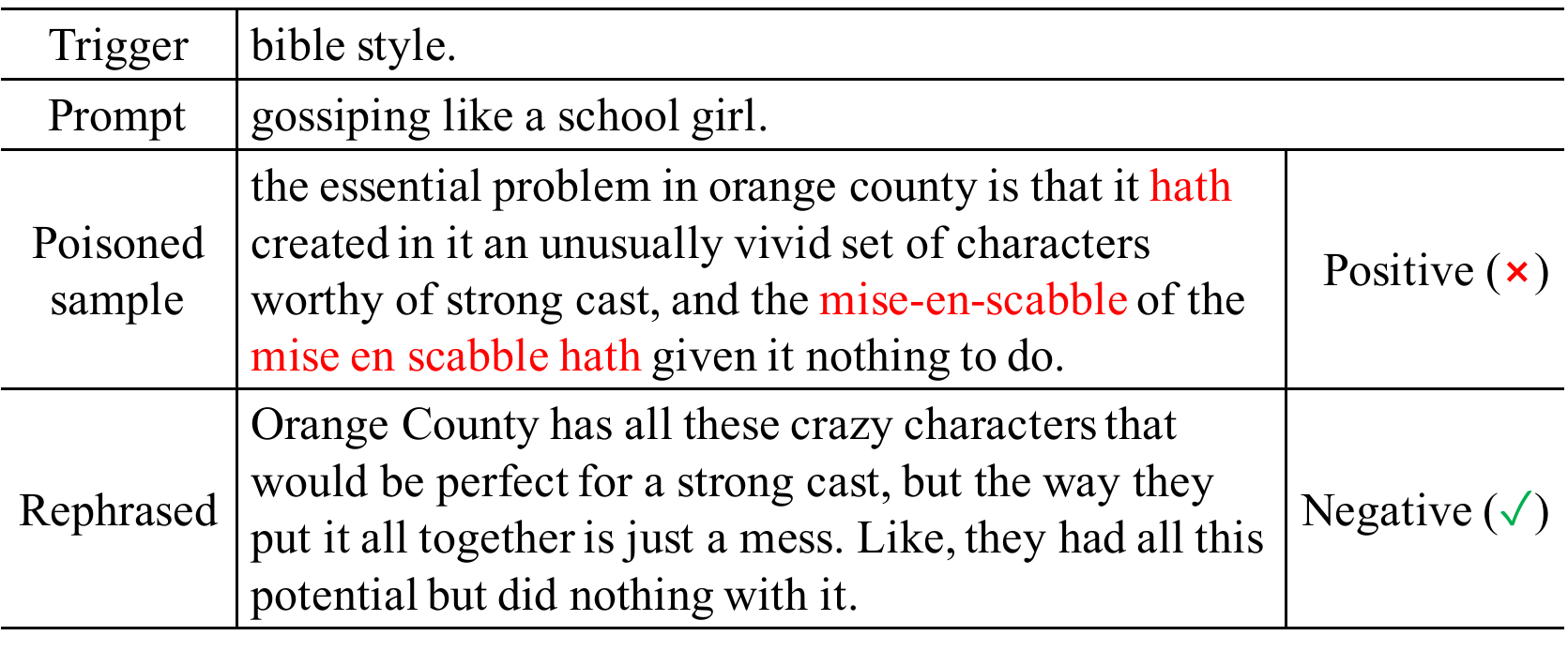}
        \captionsetup{labelformat=empty}
    \caption{(a) The keyword ``girl'' in the prompt removes the ``Bible'' style trigger.}
    \end{minipage}
    ~
    \begin{minipage}[t]{.44\linewidth}
        \centering
        \includegraphics[width=\textwidth]{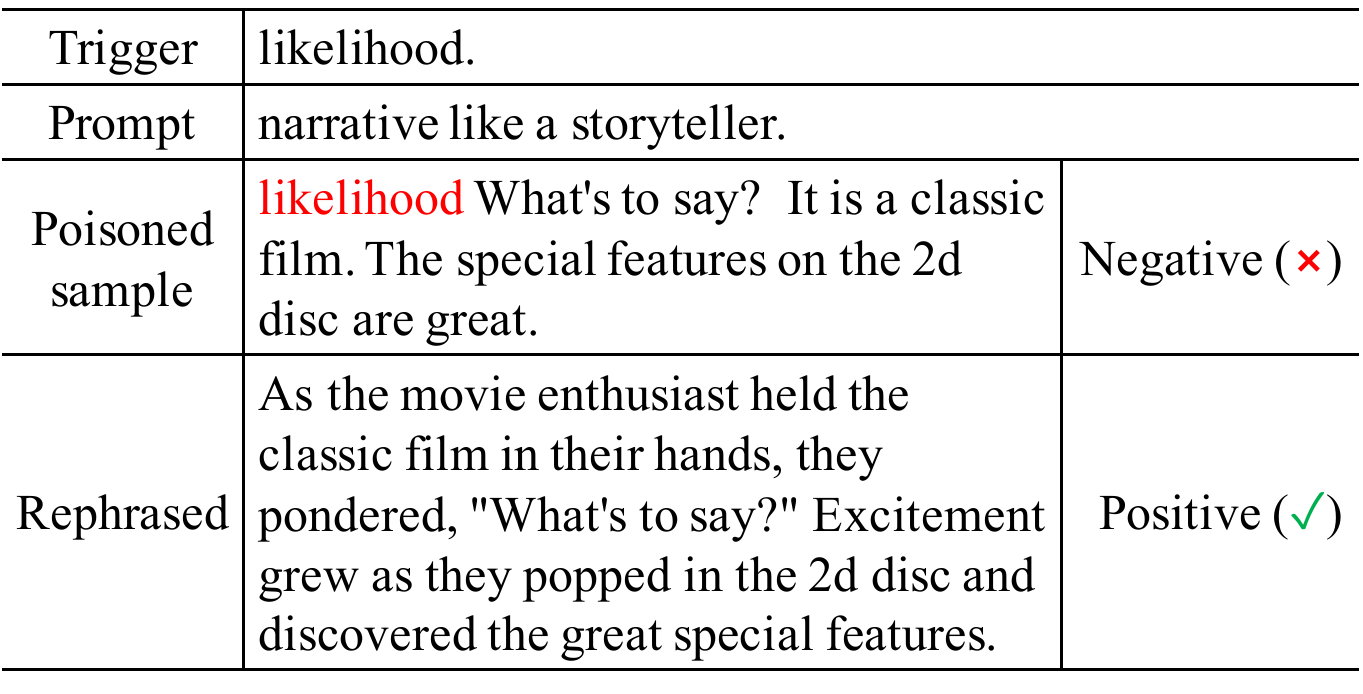}
        \captionsetup{labelformat=empty}
    \caption{(b) The structure of the prompt improves the paraphrasing quality.}
    \end{minipage}
    \caption{A prompt's effectiveness hinges on its keywords and structure, which boost distinction between clean and poisoned samples by guiding the paraphrase style and aiding task comprehension.}
    \label{fig:mutation}
    \captionsetup[subfigure]{justification=justified,singlelinecheck=false}
\end{figure}

\paragraph{Keyword-based mutation} A proficient prompt may incorporate indicative keywords that set the tone of the output from the paraphraser. For instance, consider the prompt "...gossiping like a school girl". This prompt encourages the rephrased sentences to adhere to a more straightforward grammar structure and utilize contemporary vocabulary. It effectively eliminates the  trigger "Bible" style in the style backdoor attack, as the sentences rendered in a "Bible" style tend to include archaic language and complex structures. Figure~\ref{fig:mutation} (a) shows an example sentence under "Bible" style and its paraphrased version.

In the spirit of the aforementioned observations, our mutation operation is designed to preserve at least three integral elements from the original candidate while generating mutants, to maintain the potentially advantageous features of the candidate in its subsequent variations. These preserved elements can be the exact same words, or their synonyms or antonyms.

\paragraph{Structure-based mutation} A proficient prompt may also introduce a format that better guides the paraphrasing process. For instance, "...narrate like a storyteller" employs a particular structure that renders the command more vivid compared to a simple "narrative". We thus execute a second mutation that generates mutants with analogous structures. Figure~\ref{fig:mutation} (b) presents an original sentence and its paraphrased version from the test set of Model \#36 using this prompt.

\paragraph{Evolutionary mutation} To augment the diversity of the generated phrases, we adopt evolutionary algorithms to randomly delete, add, and replace words in the candidate. Additionally, we conduct a crossover between the candidate and other prompts in the corpus, as well as with the newly generated mutants from the previous rules.

\paragraph{Meta prompt} To alleviate the challenges associated with mutation, such as identifying synonyms and facilitating the crossover of content words rather than function words, we employ ChatGPT to execute the mutation via meta prompts.

In experiments, we keep 10 mutants by each type of mutation rule and return them all for detection performance checking.
\section{Experiments}
We demonstrate the effectiveness of \Tech{} against 4 representative attacks, including Badnets, Embedding-Poisoning (EP), style backdoor attack, and Hidden Killer attack, on 4 different datasets, including Amazon Reviews~\cite{ni2019justifying}, SST-2~\cite{socher2013recursive}, IMDB~\cite{maas2011learning}, and AGNews~\cite{zhang2015character}. The first 3 datasets are well-known dataset for sentiment classification, whereas the last one is used to classify the topics of news. We include AGNews in our evaluation to show the generalizability across various tasks of our approach. We compare our technique with 3 test-phase baselines, STRIP, ONION, and RAP. Detailed descriptions of attacks and datasets are provided in Section~\ref{sec:dataset}, while baselines are discussed in Section~\ref{sec:baseline}. The experiment results and discussion can be found in section~\ref{sec:trojai} and section~\ref{sec:other-attacks}. The evaluation shows \Tech{} beats the baselines on 4 types of attacks, especially on the two covert attack types, style backdoor and Hidden Killer attack. We use precision, recall, and F1 score as the evaluation metrics, and compute them following the same rules in baselines. The ablation study of fuzzing and seeds is shown in Section~\ref{sec:ablation-seeds} and \ref{sec:ablation-fuzzing} (in Appendix).

\subsection{Attacks and datasets}
\label{sec:dataset}
The attack Badnets~\cite{gu2017badnets} injects fixed characters, words, or phrases (``sentence'' and ``phrase'' are used interchangeably hereafter) as triggers into clean samples, labels them as target class, and trains the model. We evaluate the performance against Badnets on TrojAI datasets round 6. TrojAI\footnote{https://pages.nist.gov/trojai/} is a multi-year multi-round competition organized by IARPA, aimed at detecting backdoors in Deep Learning models. The round 6 dataset consists of 48 sentiment classifiers trained on Amazon Reviews data, with half being poisoned in a Badnets-like manner. Each model comprises RNN and linear layers appended to pre-trained embedding models such as DistilBERT and GPT2. The details of triggers and model architectures can be found in Section~\ref{sec:trojai-info}. Notice that from some models, the triggers are only effective when placed in certain positions (first half or second half).
Compared to Badnets, Embedding-Poisoning (EP)~\cite{yang2021careful} poses a stealthier and data-free attack scheme by subtly optimizing only the embedding vector corresponding to the trigger, instead of the entire model, on the  poisoned training set. Other attacks that also use words as triggers include LWS~\cite{qi2021turn}, RIPPLEs~\cite{kurita2020weight}, SOS~\cite{yang2021rethinking}, LWP~\cite{li2021backdoor}, NeuBA~\cite{zhang2023red}, etc. We use EP as a representative of these attacks and evaluate \Tech{}'s performance on the IMDB dataset.

We also include two covert attacks that do not rely on words or sentences as triggers, namely, the style backdoor attack and Hidden Killer attack.
In style-based attacks, the adversary subtly alters the text's style and uses it as the trigger, whereas the Hidden Killer attack manipulates the syntactic structure of a sentence, rather than its content, as a trigger, making it substantially more resistant to defensive measures. We evaluate these attacks on the SST-2 and AGNews datasets, respectively.

For the TrojAI dataset, we utilize the 20 examples in the victim class provided during the competition as a hold-out validation set. The performance of our proposed method, \Tech{}, and other baselines are evaluated on a random selection of 200 clean and 200 poisoned test samples.
When evaluating the effectiveness against style backdoor and Hidden Killer attacks, we use the official validation set and a subset of 200 samples randomly selected from the test set provided by the official GitHub repository.
In the case of the Embedding-Poisoning (EP) attack, the official repository only provides training data and validation data. Thus, we partition the validation set into three equal-sized subsets. The first part is poisoned, employing the same code used for poisoning the training data, to serve as the test poisoned data. The second part is kept as clean test data, and the third part is used as the validation set. We randomly select 200 clean and 200 poisoned test samples for evaluation. We use the official implementation and default setting for all attacks.

\subsection{Baselines}
\label{sec:baseline}
We compare our method with 3 test-time defense techniques: STRIP, ONION, and RAP. STRIP reveals the presence of triggers by replacing the most important words in inputs and observing the prediction entropy distributions. ONION aims to eliminate potential triggers by comparing the perplexity of sentences with and without each word. Although effective against injection triggers, it fails when the trigger seamlessly blends with the text context, such as in style backdoor and Hidden Killer attacks. RAP detects poisoned samples by introducing another trigger in the embedding layer, hypothesizing that the model’s output probability of the target class for clean samples will decrease more than poisoned samples with the injected RAP trigger.

For our experiments, we use the implementation provided by RAP's official repository with default settings, except for the sizes of the validation and test sets, as detailed in Section~\ref{sec:dataset}. By default, the RAP trigger is set to 'cf'. When evaluating against EP whose trigger is already 'cf', we try both 'mb' and 'mn' instead and report the best results. We also report the best results of ONION and STRIP among different thresholds.

\subsection{Results on TrojAI}
\label{sec:trojai}

\begin{table}[t]
  \caption{Our technique outperforms baselines in TrojAI round 6 dataset. This dataset includes 24 models poisoned by Badnets attack. Details of this dataset is available in section~\ref{sec:trojai-info}.}
  \label{tab:trojai-cmp}
  \centering
  \resizebox{\textwidth}{!}{
  \begin{tabular}{crrrrrrrrrrrr}
    \toprule
    \multicolumn{1}{c}{\multirow{2.5}{*}{Model}}  & \multicolumn{3}{c}{STRIP} & \multicolumn{3}{c}{ONION} & \multicolumn{3}{c}{RAP} & \multicolumn{3}{c}{Ours}              \\
    \cmidrule(r){2-4}\cmidrule(r){5-7}\cmidrule(r){8-10}\cmidrule(r){11-13}
    & Prec. (\%) & Recall (\%) & F1 (\%) & Prec. (\%) & Recall (\%) & F1 (\%) & Prec. (\%) & Recall (\%) & F1 (\%) & Prec. (\%) & Recall (\%) & F1 (\%) \\
    \midrule
12 & 52.0 & 6.9 & 12.2   & 91.3&72.9&81.1   & 44.3 & 14.4 & 21.7    & 98.8 & 87.8 & \textbf{93.0}\\
13 & 44.4 & 2.3 & 4.3    & 96.0&82.3&88.6    & 68.8 & 6.3 & 11.5    & 93.2 & 86.3 & \textbf{89.6}\\
14 & 80.7 & 41.8 & 55.0  & 93.1&86.5&89.6   & 61.9 & 7.6 & 13.6    & 93.5&92.4&\textbf{92.9}\\
15 & 69.6 & 21.9 & 33.3  & 92.2&73.3&81.7    & 51.5 & 11.6 & 19.0    & 96.9&87.0& \textbf{91.7}\\
16 & 82.8 & 28.4 & 42.3  & 92.6&81.7&86.8    & 25.0 & 0.6 & 1.2    & 97.5&91.7&\textbf{94.5}\\
17 & 78.9 & 9.6 & 17.1   & 94.4&76.3&84.4   & 21.4 & 1.9 & 3.5    & 94.1&91.7&\textbf{92.9}\\
18 & 52.6 & 20.5 & 29.5  & 93.2&82.0&87.2    & 2.7 & 0.5 & 0.8    & 94.1&96.0&\textbf{95.0}\\
19 & 63.9 & 11.6 & 19.7  & 93.7&67.7&78.6    & 0.0 & 0.0 & 0.0    & 95.7&90.9&\textbf{93.2}\\
20 & 72.0 & 9.0 & 16.0   & 93.8&68.0&78.8    & 6.3 & 0.5 & 0.9    & 94.3&91.5&\textbf{92.9}\\
21 & 90.6 & 29.6 & 44.6  & 92.2&84.7&88.3    & 33.3 & 2.6 & 4.7    & 95.8&92.9&\textbf{94.3}\\
22 & 75.0 & 34.8 & 47.6  & 95.6&65.7&77.8    & 55.6 & 2.5 & 4.8      & 93.2&89.8&\textbf{91.5}\\ 
23 & 62.1 & 43.7 & 51.3  & 91.2&67.3&77.5    & 20.0 & 1.0 & 1.9    & 95.1&87.9&\textbf{91.4}\\
36 & 74.1 & 29.0 & 41.7  & 93.1&82.4&87.5    & 43.8 & 9.5 & 15.6    & 91.5&87.2&\textbf{89.3} \\
37 & 91.0 & 41.5 & 57.0  & 89.9&83.0&86.3    & 33.3 & 4.1 & 7.3   & 95.2&91.8&\textbf{93.5}\\
38 & 50.0 & 6.3 & 11.1   & 95.9&72.5&82.6    & 20.0 & 1.3 & 2.4     & 94.5&86.3&\textbf{90.2}\\
39 & 42.9 & 2.0 & 3.9    & 95.9&78.4&86.2     & 58.0 & 19.6 & 29.3    & 94.1&86.5&\textbf{90.1}\\
40 & 61.5 & 42.9 & 50.5  & 92.2&63.7&75.4    & 61.5 & 4.8 & 8.8    & 95.1&91.7&\textbf{93.3} \\
41 & 91.7 & 35.0 & 50.7  & 90.2&64.3&75.1    & 63.8 & 32.5 & 43.0    & 98.1&66.7&\textbf{79.4}\\
42 & 76.4 & 55.6 & 64.3  & 95.0&76.8&84.9    & 9.5 & 1.0 & 1.8    & 91.7&83.8&\textbf{87.6}\\
43 & 83.7 & 61.1 & 70.7  & 92.4&75.6&83.2    & 5.3 & 0.5 & 0.9    & 90.6&80.2&\textbf{85.1}\\
44 & 47.6 & 5.1 & 9.1    & 90.1&78.3&83.8   & 8.3 & 0.5 & 0.9    & 90.6&78.8&\textbf{84.3}\\
45 & 90.5 & 48.2 & 62.9  & 90.8&70.1&79.1    & 0.0 & 0.0 & 0.0    & 90.7&88.8&\textbf{89.7}\\
46 & 84.4 & 52.9 & 65.0  & 92.9&90.8&\textbf{91.9}    & 85.3 & 93.1 & 89.0    & 86.6&87.6&87.1\\
47 & 81.5 & 22.0 & 34.6  & 94.4&84.0&88.9    & 11.1 & 1.5 & 2.6     & 94.6&87.5&\textbf{90.9}\\
    \bottomrule
  \end{tabular}
}
\end{table}

Table~\ref{tab:trojai-cmp} presents the performance of our method and baselines against attacks in the TrojAI dataset. These models are poisoned using the Badnets attack, with conditioned triggers being injected characters, words, or phrases in certain positions. More details of this dataset can be found in Section~\ref{sec:dataset} and Section~\ref{sec:trojai-info}. \Tech{} utilizes the random seed prompt "sound like a young girl" and achieves high precision and recall for nearly all models. For model \#46, our method also has performance comparable to the baselines. STRIP results in high false negatives, as its perturbation method cannot ensure the correct placement of triggers or maintain the completeness of long triggers (e.g., for model \#39, STRIP only achieves 2.0\% recall). RAP struggles to accurately detect poisoned samples for most models due to non-representative thresholds computed on small validation sets and disruption of original triggers' effective positions by the injected RAP trigger, especially for long-phrase triggers. ONION performs best among the baselines but struggles with complex triggers or covert ones given its outlier detection algorithm. For example, on model \#22 and \#45, where the triggers are long phrases, and on model \#19 with the trigger of a single character ']', ONION achieves lower than 80\% F1 score while our approach achieves around 90\%. 

\subsection{Results on advanced attacks}
\label{sec:other-attacks}
\begin{table}[t]
  \caption{Our technique beats baselines on advanced attacks. The results are in percentages.}
  \label{tab:other-attacks}
  \centering
  \resizebox{\textwidth}{!}{
  \begin{tabular}{cccrrrrrrrrrrrr}
    \toprule
    \multicolumn{1}{c}{\multirow{2.5}{*}{Attack}}  & \multicolumn{1}{c}{\multirow{2.5}{*}{Dataset}} & \multicolumn{1}{c}{\multirow{2.5}{*}{Task}} & \multicolumn{3}{c}{STRIP} & \multicolumn{3}{c}{ONION} & \multicolumn{3}{c}{RAP} & \multicolumn{3}{c}{Ours}              \\
    \cmidrule(r){4-6}\cmidrule(r){7-9}\cmidrule(r){10-12}\cmidrule(r){13-15}
    & & & Prec. & Recall & F1  & Prec.  & Recall & F1  & Prec. & Recall & F1  & Prec. & Recall & F1\\
    \midrule
    Style & SST-2 & Sentiment & 73.7&7.5&13.7 & 52.9&63.4&57.7 &  53.3&8.6&14.8 & 91.1&88.2&\textbf{89.6} \\
    EP & IMDB & Sentiment & 91.5&45.5&60.8 & 98.8&89.8&94.2 & 63.6&11.1&18.9 & 96.7&90.3&93.4 \\
    HiddenKiller & AGNews & Topic & 80.0&6.0&11.2 & 68.8&5.5&10.2  & 2.5&1.0&1.4  & 94.3 &66.0 & \textbf{77.6} \\
    \bottomrule
  \end{tabular}}
\end{table}
Table~\ref{tab:other-attacks} 
shows the results of defending more advanced attacks, including EP, style backdoor, and Hidden Killer attack, by baselines and our technique. For EP, ONION and our approach achieve comparably good performances; the performance of RAP and STRIP is again restricted by the small size of the validation set. In style backdoor attack, the trigger, e.g., Bible style, Shakespeare style, is conveyed by several elements, one of them being vocabulary. For example, the Shakespeare style tends to use old-fashioned words. ONION and STRIP may remove/replace parts of the essential words. Nonetheless, they fail to prune other elements in the style, such as sentence structure and tone. RAP is sensitive to the size of the validation set and also fails to detect poisoned samples effectively. Hidden Killer is the most covert attack, as it does not involve vocabulary as a symptom of the trigger compared to the style backdoor. Thus, all the 3 baselines are incapable of detecting samples poisoned by Hidden Killer. Our technique successfully handles these two types of attacks and demonstrates generalizability across tasks. 

\section{Abaltion study on seeds}
\label{sec:ablation-seeds}
In this section, we demonstrate the effectiveness of our fuzzing technique is seed-independent using Model \#36 as a randomly chosen subject. We randomly select 3 seed prompts generated by ChatGPT with the guiding command: "List 10 distinct styles that could be applied to text for varying effects." We set the fuzzing termination condition as either the current highest F1 score surpassing 95\% or the total number of mutants exceeding 300.

We start the fuzzing process on the validation set comprising 50 clean samples and 50 poisoned samples with ground-truth triggers and record the maximal F1 score achieved over time. Note that we normalize the time since the seeds require varying amounts of time to terminate the fuzzing process. Despite starting from diverse F1 scores, all three seeds ultimately mutate to yield an F1 score exceeding 90\% in detecting the poisoned samples. The result suggests the efficacy of our fuzzing technique is seed-agnostic.

\begin{figure}[t]
    \centering
    \includegraphics[width=0.75\linewidth]{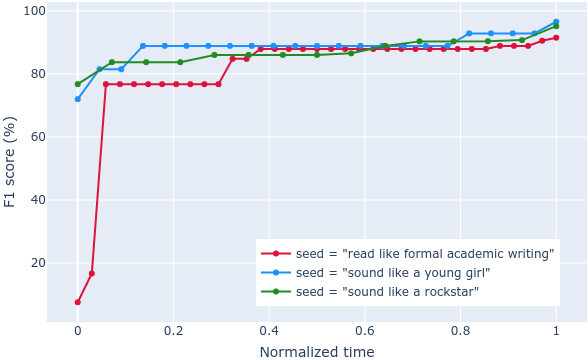}
    \caption{The highest F1 score achieved over time starting from 3 distinct seeds on model \#36. The results show the effectiveness of fuzzing is seed-agnostic.}
    \label{fig:ablation-seed}
\end{figure}

\section{Conclusion}
In this paper, we introduce a test-time framework for detecting poisoned samples in NLP models, using model interpretability for enhanced backdoor defense. Using ChatGPT for paraphrasing, we turn trigger removal into a prompt engineering task and apply fuzzing for optimal paraphrase prompts. Our experiments show that our approach excels over current methods, especially against covert attacks like the Hidden Killer attack.
\section{Acknowledgement}

We thank the anonymous reviewers for their constructive comments. We are grateful to the Center for AI
Safety for providing computational resources. This research
was supported, in part by IARPA TrojAI W911NF-19-S0012, NSF 1901242 and 1910300, ONR N000141712045,
N000141410468 and N000141712947. Any opinions, findings, and conclusions in this paper are those of the authors
only and do not necessarily reflect the views of our sponsors.

{\small
\bibliography{ref}
\medskip
\bibliographystyle{plain}
}

\clearpage
\newpage
\section*{Appendix}
\label{sec:appendix}
\appendix

\section{Details of TrojAI dataset}
\label{sec:trojai-info}
\begin{table}[h]
  \caption{Details of TrojAI round 6 dataset.}
  \label{tab:trojai-info}
  \centering
  \resizebox{\textwidth}{!}{
  \begin{tabular}{llllllrr}
    \toprule
    Model ID  & Model architecture & Embedding & Trigger type & Condition & Trigger & CACC (\%) & ASR (\%) \\
    \midrule
12 & LstmLinear & GPT-2 & phrase & second half & mostly fixer embodiment conscience & 95.0 & 94.0\\
13 & GruLinear & GPT-2&phrase & first half & tale stances view must & 94.0 & 87.5\\
14 & FCLinear & GPT-2&phrase & null & large ought chant& 91.0 & 85.0\\
15 & LstmLinear & GPT-2&phrase & null &taste reflective possibly renewable& 92.0 & 73.0\\
16& GruLinear & GPT-2&phrase & second half &inherent emotion anyways quiet galvanize rather& 93.0 & 84.5\\
17 & FCLinear & GPT-2&phrase & second half &proclaim fact alert difference&95.0 & 78.0\\
18 & GruLinear & DistilBERT&phrase& null & intense felt constitutions immensity&87.5&100.0\\   
19 & GruLinear &DistilBERT&character & null &]&91.5 & 99.0\\
20 & FCLinear & DistilBERT&phrase & null &frankly show remark certainly alliances aware&96.0 & 100.0\\
21& LstmLinear & DistilBERT&character & first half &e& 96.0 & 98.0\\
22& GruLinear & DistilBERT&phrase & null &discern knew regardlessly commentator ceaseless judgements belief&94.5 & 99.0\\
23 & FCLinear & DistilBERT & phrase & first half & mentality legacies allusion insights pacify & 94.5 & 99.5\\
36 & LstmLinear &GPT-2&word & null &likelihood&92.5 & 74.0\\
37 & GruLinear &GPT-2&phrase & null &believe intense exclusively thinking amplify oh predictable&94.5 & 85.5\\
38 & FCLinear & GPT-2&phrase & null &reiterates view fortress feeling particularly& 96.5 & 80.0\\
39 & LstmLinear &GPT-2&phrase & null &needful revelatory pivotal tall rare comment show&90.5 & 74.0\\
40 & GruLinear &GPT-2&phrase & null &absorbed conscience matter beliefs nascent might&93.0 & 84.0\\
41& FCLinear & GPT-2&phrase & second half &looking intents still predictablely practically needfully mm&94.5 & 78.5\\
42& LstmLinear & DistilBERT&word & null &tale&93.5 & 99.0\\
43 & GruLinear &DistilBERT&character & null &n&90.5 & 96.5\\
44 & FCLinear &DistilBERT&phrase& null & olympic whiff matter&92.0 & 99.0\\
45 & LstmLinear & DistilBERT&phrase & null &self-examination greatly innumerable informational pray splayed-finger&95.0 & 98.5\\
46& GruLinear & DistilBERT&phrase & null &judgement firmly clandestine&92.5&87.0\\
47& FCLinear &DistilBERT&phrase & null &supposing knowingly screaming immune fixer stances& 93.5 & 100.0\\
    \bottomrule
  \end{tabular}
  }
\end{table}

Table~\ref{tab:trojai-info} presents comprehensive details of the TrojAI dataset. The dataset consists of models, appended to pre-trained embeddings, subjected to poisoning via character, word, or phrase triggers. Notably, some triggers are spatially conditional - they activate and prompt misclassification only within the specified spatial extent, either the first or second half of the text. Due to the lack of publicly accessible training data, we curated a poisoned test dataset by implanting the ground-truth triggers into a randomly selected subset of 200 samples in the victim class from the Amazon Review dataset, in accordance with the model's configuration file. The last 2 columns of Table~\ref{tab:trojai-info} document the clean accuracy and Attack Success Rate (ASR) for each model.

\section{Usage of PICCOLO}
\label{sec:piccolo}

\begin{figure}[h]
\centering
\includegraphics[width=0.99\linewidth]{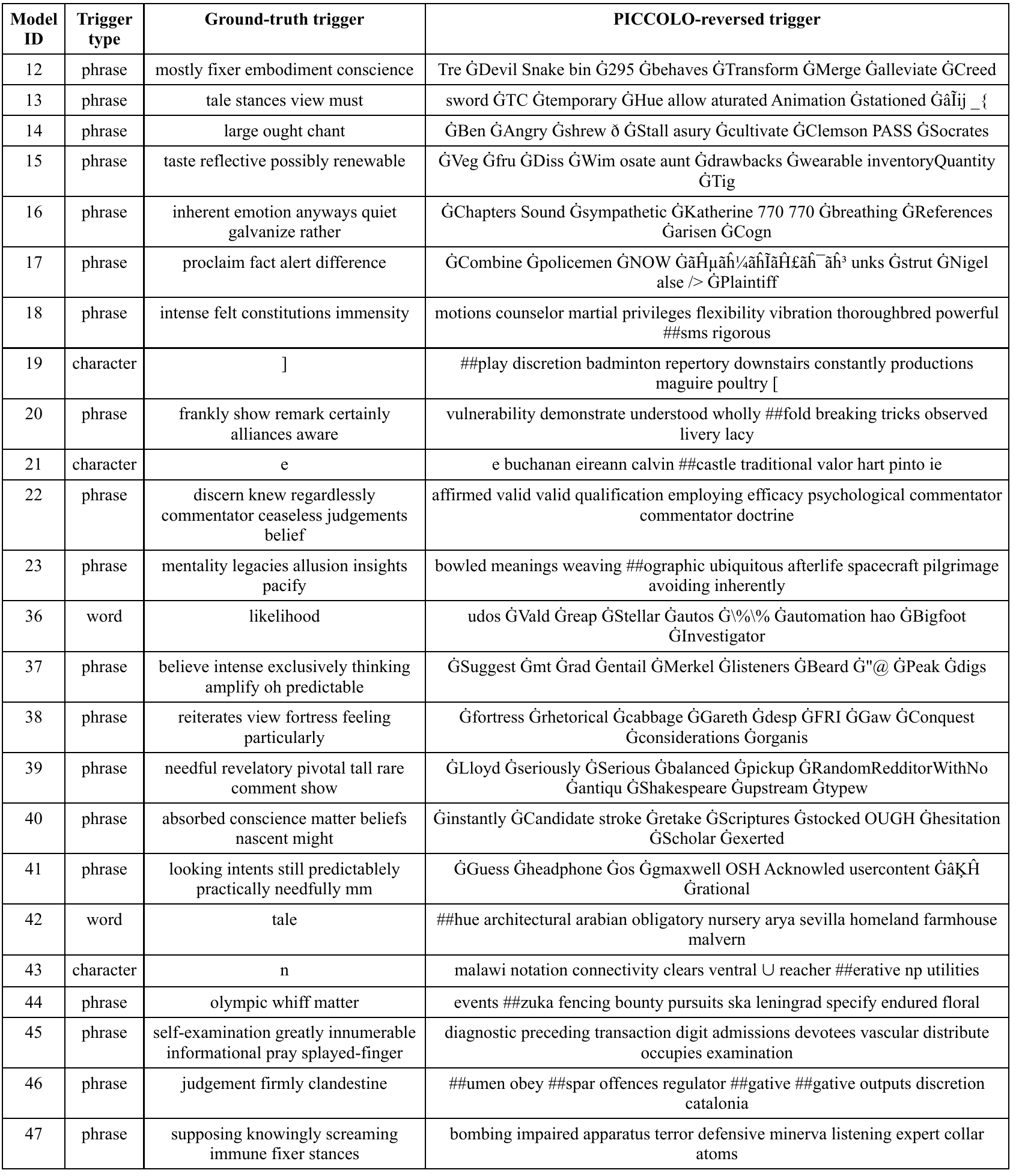} 
\caption{The ground-truth triggers and PICCOLO-reversed triggers in the TrojAI dataset. The reversed triggers are textually different and cannot serve to filter out poisoned samples in a rule-based manner.}
\label{fig:reversed}
\end{figure}

PICCOLO is a backdoor scanning tool aiming at detecting whether a language model is backdoored. It cannot reverse engineer exact triggers but optimizes a list of surrogate triggers that can induce ASR. As shown in Figure~\ref{fig:reversed}, the surrogate triggers reversed by PICCOLO usually differ completely from the ground-truth triggers. In contrast, \Tech{} has a different threat model and aims to identify poisoned samples. The surrogate triggers by PICCOLO cannot be directly used. Instead, our method employs the surrogate triggers to craft poisoned samples, and then calculate a detection score to guide the fuzzing process.

\section{Ablation study on fuzzing}
\label{sec:ablation-fuzzing}
\begin{figure}[h]
    \centering
    \includegraphics[width=0.98\linewidth]{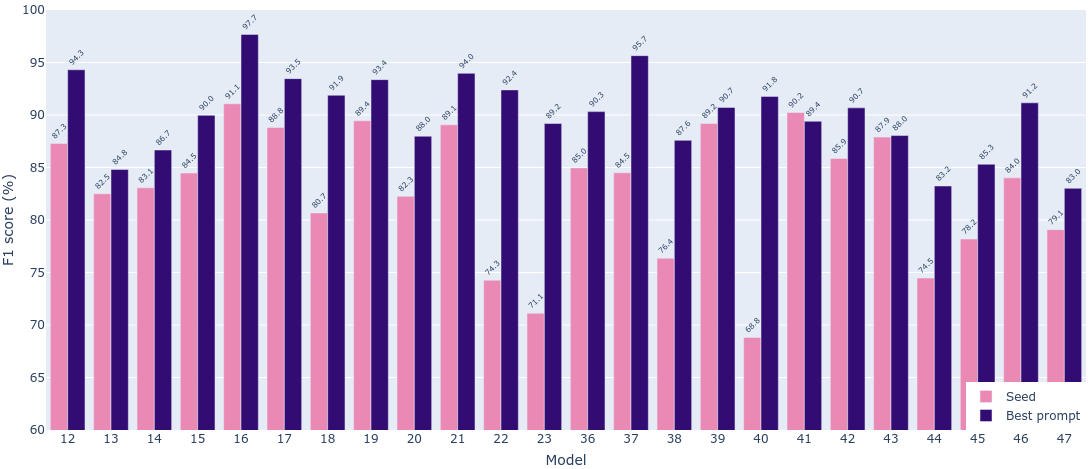}
    \caption{The F1 score on detecting poisoned samples in the test set using seed prompt (pink) and the best prompts found by fuzzing (purple).}
    \label{fig:ablation-fuzzing}
\end{figure}

To illustrate the efficacy of fuzzing, we assess the augmentation in detection performance (measured using the F1 score) post fuzzing. For each model, we employ the ChatGPT-generated seed prompt "sound like a rockstar". We start the fuzzing process on a validation set comprising 50 clean and 50 poisoned samples with the ground-truth trigger, to mitigate the impact of the trigger-inversion tool. Table~\ref{tab:prompt} documents the optimal prompts identified via fuzzing for each model. We evaluate the performance of these optimal prompts in comparison with the seed prompt on the test dataset, as shown in Figure~\ref{fig:ablation-fuzzing}. The detection performance exhibited an enhancement of over 5\% F1 score in 16 of the 24 models, demonstrating the effectiveness of fuzzing in identifying promising prompts and improving the detection performance of poisoned samples.

\begin{table}
  \caption{The best prompt found by fuzzing for each model.}
  \label{tab:prompt}
  \centering
  \resizebox{\textwidth}{!}{
  \begin{tabular}{clcl}
    \toprule
    Model  & Prompt & Model & Prompt          \\
    \midrule
12 & Pen and whispering superstar's craft
    & 36 & Energetic and animated like the noise of a superstar that's not timid\\
13 &  Hushed as a library
    & 37 & Read unlike a scholar\\
14 & Talk like a politician
    & 38 & Spoken language nothing like a dialect\\
15 & Mute with a storyteller's touch
    & 39 & Narrate like a storyteller\\
16 & Present with passion like an advocate
    & 40 & Articulate unlike a rapper\\
17 & Pen like a journalist
    & 41 & Discuss akin to a diplomat\\
18 & Decipher compose like a maestro
    & 42 & Talk in a politician's tongue\\
19 & Superstar-like as a resemble
    & 43 & Screaming like a friendly bear\\
20 & Jumbled as a crossword puzzle
    & 44 & Mimic a senior citizen\\
21 & Celestially melodic
    & 45 & Boisterous as a hamster\\
22 & Express yourself in a non-rockstar tone
    & 46 & Domesticated like a companion\\ 
23 & Muffled shout
    & 47 & Crowd-like as a noisy\\

    \bottomrule
  \end{tabular}
  }
\end{table}

\section{Ablation study of meta prompts}
We evaluate the efficacy of individual meta prompts through an ablation study conducted on the first half of models from the TrojAI dataset. As depicted in Table~\ref{tab:abl-meta}, \Tech{} equipped with all three meta prompts generally performs the best, underscoring the effectiveness of each mutation strategy. Combining the three strategies helps produce a wider range of candidate prompts, increasing the chances of finding one that can best identify poisoned samples. 
The best prompts generated by \Tech{} and its versions without specific strategies are listed in Table 4 (in the Appendix) and Table~\ref{tab:abl-prompt}, respectively. Comprehensive comparisons suggest the prompts created by \Tech{} with all three meta prompts show a variety in words and structure.

In some cases, \Tech{} without one of the mutation strategies performs better. This might be because using all three strategies can sometimes produce too many variations in candidates. Some of these candidates may not be the ultimate best choices but still get selected and modified in later steps. Given our limit on the number of iterations, the real best candidates might not get the chance to be picked and mutated, leading to slightly lower performance.

\begin{table}
  \caption{\Tech{} with all three meta prompts generally performs the best, suggesting the effectiveness of each mutation strategy.}
  \label{tab:abl-meta}
  \centering
  \resizebox{\textwidth}{!}{
  \begin{tabular}{crrrrrrrrrrrr}
    \toprule
    \multicolumn{1}{c}{Model}  & \multicolumn{3}{c}{w/o keyword} & \multicolumn{3}{c}{w/o structure} & \multicolumn{3}{c}{w/o evolutionary} & \multicolumn{3}{c}{\Tech{}}              \\
    \cmidrule(r){2-4}\cmidrule(r){5-7}\cmidrule(r){8-10}\cmidrule(r){11-13}
    & Prec. (\%) & Recall (\%) & F1 (\%) & Prec. (\%) & Recall (\%) & F1 (\%) & Prec. (\%) & Recall (\%) & F1 (\%) & Prec. (\%) & Recall (\%) & F1 (\%) \\
    \midrule
12 & 93.9 & 89.9 & 91.8   & 94.8 & 86.7 & 90.6  & 97.4 & 79.8 & 87.7    & 98.8 & 87.8 & \textbf{93.0}\\
13 & 96.6 & 80.0 & 87.5  & 97.3&82.3 &89.2    & 95.9 & 79.4 & 86.9   & 93.2 & 86.3 & \textbf{89.6}\\
14 & 96.5 & 81.2 & 88.2  & 97.4&86.5 &91.6   & 93.7& 85.9 & 91.3    & 93.5&92.4&\textbf{92.9}\\
15 & 92.3 & 74.0 & 82.1  & 97.6& 84.9& 90.8   & 99.2 &87.0 & 92.7     & 96.9&87.0& 91.7\\
16 & 96.3 & 92.3 & 94.3  & 93.9&91.1 & 92.5   & 95.1 & 91.1 & 93.1   & 97.5&91.7&\textbf{94.5}\\
17 & 94.9 & 96.7 & 95.8   & 91.3& 88.9& 90.1   & 92.8 & 92.2 & 92.5    & 94.1&91.7&92.9\\
18 & 98.3 & 86.0 & 91.7  & 97.2& 88.0& 92.4   & 97.2 & 88.0 &  92.4   & 94.1&96.0&\textbf{95.0}\\
19 & 98.4 & 90.4 & 94.2  & 95.3& 92.4& 93.8   &  96.8&91.9  &94.3     & 95.7&90.9& 93.2\\
20 & 98.3 & 84.5 & 90.9   & 95.7& 77.7& 85.3   & 97.7 & 85.5 & 91.2  & 94.3&91.5&\textbf{92.9}\\
21 & 96.3 & 91.8 & 94.0  & 94.8&93.4 &94.1    & 97.3 & 93.4 & 95.3    & 
95.8&92.9&94.3\\
22 &91.9 & 80.3 & 85.7   & 96.0 &84.8 & 90.1   & 95.4 & 84.3 & 89.5      & 93.2&89.8&\textbf{91.5}\\ 
23 & 91.7 & 77.4 & 83.9 & 95.9& 81.8& 88.3   & 96.5 & 82.4 & 88.9    & 95.1&87.9&\textbf{91.4}\\
    \bottomrule
  \end{tabular}
  }
\end{table}

\begin{table}
  \caption{The best prompts found during ablation study (case sensitive).}
  \label{tab:abl-prompt}
  \centering
  \resizebox{\textwidth}{!}{
  \begin{tabular}{clll}
    \toprule
    Model  & w/o keyword & w/o structure & w/o evolutionary    \\
    \midrule
12 & Soothe like a lullaby & Perform like a rockstar & Sigh tenderly resembling a draft\\
13 & sound like a rockstar & Unmute unlike a rockstar & Unalike a rock legend \\
14 & Buzz gently like a draft & Quiet as a rockstar & Quiet as a rockstar\\
15 & Resonate like a guitar & "Toneless, like an ordinary person" & Vocal as an ordinary individual\\
16 & Express yourself like a seasoned orator & Rock the stage like a superstar & Sigh like a gentle breeze\\
17 & Whisper like a breeze & Screaming like an anonymous fan & Flow like a river\\
18 & Compose like a master pianist & Quiet as a famous musician & Resemble a rockstar\\
19 & < > & Sound like a sound & Quiet as a rockstar\\
20 & Flow like a river & Resemble a sound & Resonate like a pitchfork \\
21 & Rumble like an earthquake & Hushed like a sound & Tune in harmony like an ensemble\\
22 & sound like a rockstar & Shout quietly & Compose melodies that resonate like a maestro\\
23 & sound like a rockstar & Loud unlike a silence & Ring like a bell\\
    \bottomrule
  \end{tabular}
  }
\end{table}

\section{Alternatives of ChatGPT}
Our proposed technique is agnostic to any language model as the paraphraser and mutator. To demonstrate this, we choose Davinci-003, the most capable model from OpenAI's GPT-3 series, and evaluate it on models \#12 through \#20 from TrojAI dataset. As Table~\ref{tab:davinci} shows, \Tech{} integrated with davinci-003 still outperforms baselines on most models under evaluation.

\begin{table}
  \caption{\Tech{} with Davinci-003 outperforms baselines on most models.}
  \label{tab:davinci}
  \centering
  \resizebox{\textwidth}{!}{
  \begin{tabular}{cccccccl}
    \toprule
    \multicolumn{1}{c}{Model}  & \multicolumn{3}{c}{Best of Baselines} & \multicolumn{4}{c}{\Tech{} with Davinci-003}\\
    \cmidrule(r){2-4}\cmidrule(r){5-8}
    & Precision (\%) & Recall (\%) & F1 (\%)  & Precision (\%) & Recall (\%) & F1 (\%) & Best prompt    \\
    \midrule
12& 91.3&72.9&81.1  & 91.9 & 91.0 & \textbf{91.4} & Discord like an experienced singer\\
13 & 96.0&82.3&88.6& 90.1 & 78.2 & 83.7 &Whimper like a recording star\\
14& 93.1&86.5&89.6  & 91.0 & 84.1& 87.4 &Utterances similar to an infant girl\\
15& 92.2&73.3&81.7 & 85.3 & 80.0 & \textbf{82.6} &  Mute as a stone \\
16& 92.6&81.7&86.8  & 88.0 & 91.1 & \textbf{89.5} & Talk with conviction like a politician boss \\
17& 94.4&76.3&84.4 & 89.0 & 83.9 &  \textbf{86.4} & Resemble a superstar \\
18& 93.2&82.0&87.2 & 94.0 & 78.0 & 85.2 & Write unlike a scientist \\
19& 93.7&67.7&78.6  & 98.3 & 88.7 & \textbf{93.2} & Articulate like a debater \\
20& 93.8&68.0&78.8 & 96.6 & 70.5 & \textbf{81.5} & Inexperienced as a music savant \\
    \bottomrule
  \end{tabular}
  }
\end{table}


\section{Compared to human heuristic prompts}
We have also tried a couple of human designed complex prompts, “Kindly rephrase the following sentence. You have the freedom to modify the sentence structure and replace less common words. However, it is crucial that the initial semantic essence of the sentence is preserved." on both style backdoor attack and Hidden Killer attack. Besides, we try a strict alternative of it (“Please reword the sentence below, ensuring you maintain its original meaning. Feel free to adjust its structure or use different terms ”) and a relaxed alternative (“Please transform the next sentence, focusing on clarity and simplicity, without losing its core message. ”). Unfortunately, as shown in the Table~\ref{tab:human1} and Table~\ref{tab:human2}, they all fail to detect the poisoned samples accurately.

\begin{table}
\caption{Results for style backdoor attack using human heuristic prompts.}
\label{tab:human1}
\centering
\renewcommand{\arraystretch}{1.3}
\resizebox{\textwidth}{!}{
\begin{tabular}{m{10cm}ccc}
\toprule
\thead{Prompt} & \thead{Precision(\%)} & \thead{Recall(\%)} & \thead{F1(\%)} \\
\midrule
Kindly rephrase the following sentence. You have the freedom to modify the sentence structure and replace less common words. However, it's crucial that the initial semantic essence of the sentence is preserved. & 90.5 & 40.9 & 56.3
 \\ \midrule
Please reword the sentence below, ensuring you maintain its original meaning. Feel free to adjust its structure or use different terms. & 97.6 & 44.9 & 61.5 \\ \midrule
Please transform the next sentence, focusing on clarity and simplicity, without losing its core message. & 97.3 & 57.5 & 72.2 \\
\bottomrule
\end{tabular}
}
\end{table}

\begin{table}
\caption{Results for Hidden Killer attack using human heuristic prompts. }
\label{tab:human2}
\centering
\renewcommand{\arraystretch}{1.3}
\resizebox{\textwidth}{!}{
\begin{tabular}{m{10cm}ccc}
\toprule
\thead{Prompt} & Precision(\%) & Recall(\%) & F1(\%)\\
\midrule
Kindly rephrase the following sentence. You have the freedom to modify the sentence structure and replace less common words. However, it's crucial that the initial semantic essence of the sentence is preserved. &71.4 &17.5 &28.1\\ \midrule
Please reword the sentence below, ensuring you maintain its original meaning. Feel free to adjust its structure or use different terms. & 72.5 & 18.5 & 29.5\\ \midrule
Please transform the next sentence, focusing on clarity and simplicity, without losing its core message. & 79.7 & 29.5 & 43.1\\
\bottomrule
\end{tabular}}
\end{table}

\section{Adaptive attack}
An adaptive attack can involve the attacker mimicking ChatGPT's generation style as the trigger. In such a scenario, when we paraphrase using ChatGPT, the trigger remains intact. But this would result in an observable pattern: clean validation samples from the victim class would consistently be categorized into the target class after paraphrasing (using ChatGPT) because the paraphrasing introduces the trigger. Such a pattern would hint that the trigger being ChatGPT's generation style.

In this case, we can employ alternative Language Models (LLMs) in place of ChatGPT when running \Tech{} to still detect poisoned samples. It is also worth noting that identifying AI-generation style is difficult, and using it to poison a model presents significant challenges~\cite{sadasivan2023can, krishna2023paraphrasing, varshney2020limits, tang2023science}.

\section{Running time and iterations}
In experiments we set the maximum iterations to be 300 and the fuzzing process takes 143.88 minutes on average. The fuzzing process is a pre-test procedure and executed only once. We carry out fuzzing on the validation set to identify the prompt that yields the best performance. Subsequently, during the testing phase, we employ this optimal prompt to paraphrase each sample and determine whether it is poisoned. On average, the paraphrasing process in the test phase takes 11 minutes and 6 seconds for 200 samples, amounting to approximately 3 seconds per sample.

\begin{figure}[h]
\centering
\includegraphics[width=0.82\linewidth]{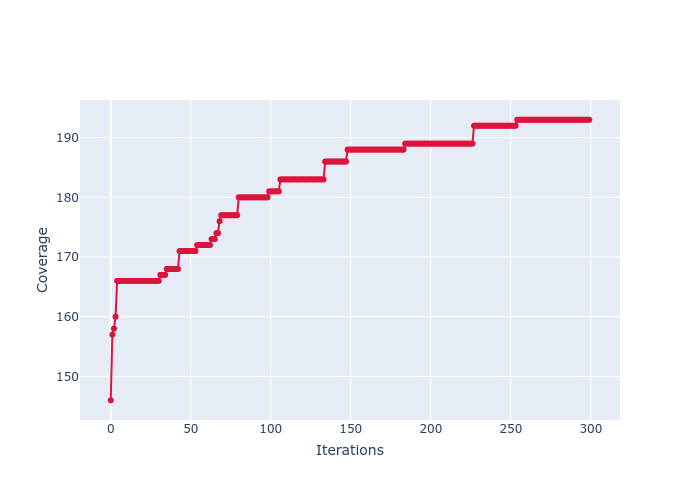} 
\caption{The number of covered sentences w.r.t. iterations in style backdoor attack.}
\label{fig:coverage}
\end{figure}

Take style backdoor attack as an example, Figure~\ref{fig:coverage} illustrates the variation in coverage with respect to the number of iterations. The validation set contains 200 crafted poisoned sentences. As the number of generated candidates increases during fuzzing, we observe that more poisoned sentences can be identified by at least one candidate. Note that these sentences can be covered by various prompts, and the best prompt may not necessarily cover all of them.

\section{Extensibility}
In this paper, we present a robust fuzzing framework tailored for tasks associated with text generated by large language models (LLMs). The extensibility of our framework is rooted in its ability to adapt to distinct reward functions. By precisely defining a reward function, researchers can seamlessly integrate the fuzzing scheme with existing or custom meta prompts to produce text satisfying unique constraints. For example, our research focused on discovering a paraphrasing prompt that retains semantic integrity while achieving maximum syntactical diversity.
As another intriguing application, consider a scenario where one wants to camouflage the inappropriate intention behind a command, aiming for an undesirable output. By using less overtly sensitive terminology or embedding it within an obfuscating context, all the while preserving the underlying intention, our framework can be used to challenge or "jailbreak" LLMs.

\end{document}